# The role of pawnshops in risk coping in early twentieth-century Japan*

Tatsuki Inoue†

**Abstract**

This study examines the role of pawnshops as a risk-coping device in prewar Japan. Using data on pawnshop loans for more than 250 municipalities and exploiting the 1918–1920 influenza pandemic as a natural experiment, we find that the adverse health shock increased the total amount of loans from pawnshops. This is because those who regularly relied on pawnshops borrowed more money from them than usual to cope with the adverse health shock, and not because the number of people who used pawnshops increased.

* I would like to express my gratitude to Tetsuji Okazaki, Kota Ogasawara, and participants at the Economic History Society Annual Conference 2019 at Queen's University Belfast and 2019 Japanese Economic Association Spring Meeting at Musashi University for their helpful comments. This work was supported by Grant-in-Aid for JSPS Fellows (Grant Number: 17J03825). Any errors are my own.

† Graduate School of Economics, The University of Tokyo, Akamon General Research Building, 3F 353, 7-3-1, Hongo, Bunkyo-ku, Tokyo 113-0033, Japan. E-mail: inoue-tatsuki245@g.ecc.u-tokyo.ac.jp.

# Introduction

Most industrialized countries were characterized by huge income inequality before World War II (Piketty 2014). Since formal systems of social insurance were underdeveloped, the poor were more vulnerable to unforeseen accidents such as illness than today's poor people in developed countries. Furthermore, an increase in migration removed people from the traditional informal social insurance systems provided by their local communities (Gorsky 1998).

Given the foregoing, many studies have examined how low-income people smoothed their consumption under earnings shocks in the late nineteenth century and early twentieth century.[1] Kiesling (1996) demonstrated that workers in Lancashire generally relied on informal sources such as savings, intrafamily transfers, and personal networks around the time of the cotton famine (1861–1865). Horrell and Oxley (2000) found that informal strategies, such as eliciting labor force participation by family members and reducing rent, were more important for industrial households in Britain at the end of the nineteenth century than benefits from formal institutions. In Helsinki in the 1920s, Saaritsa (2008) pointed out the importance of informal transfers between friends, relatives, and neighbors. In addition, Saaritsa (2011) found that informal assistance contributed to smoothing the income of worker households in 1928. James and Suto (2011) showed that Japanese workers in the 1910s and 1920s tended to have postal savings, which helped them smooth consumption. Ogasawara (2018a) used data on working class households in Osaka around 1920 to confirm that the response of borrowing to idiosyncratic shocks was greater than that of other income sources.

These studies indicate that informal sources such as savings and borrowing from acquaintances were widely used to mitigate earnings shocks. Their findings, however, are based on household data in limited areas or industries. Few studies cover the whole of a country. Hollis and Sweetman (1998) and Scott and Walker (2012) covered the whole of Ireland and Britain, respectively, but little is known about risk-coping behavior in the face of unexpected shocks. Hollis and Sweetman (1998) focused on showing that microcredit societies (i.e., loan funds) were better than banks at lending to the poor in the 1940s. Scott and Walker (2012) examined expenditure smoothing during expenditure crises of the family lifecycle rather than under exogenous shocks. To bridge this knowledge gap, we focus on pawnshops, which spread far and wide in early twentieth-century Japan. Our observations consist of over 250 municipalities (i.e., cities and counties) across Japan.

---

[1] With regard to the early modern period, Ogilvie et al. (2012) revealed that even rural people in the Black Forest borrowed for a variety of purposes including consumption smoothing.

Private pawnshops, which date back to the Kamakura period (1185–1333), were the primary financial institutions for poor people in Japan even in the 1920s, in contrast to Western countries (see Murhem 2015; Saarista 2011, p. 107; Scott and Walker 2012, p. 798). Surprisingly, they provided approximately 25 million loans per year. Indeed, pawnshops even accepted inexpensive goods, such as clothes, as pawns. Thus, even poor people could access loans. Meanwhile, people could obtain small short-term loans as needed. This feature was suitable as a risk-coping strategy.

To investigate the response of borrowing from pawnshops to an unforeseen shock, we use data on loans from pawnshops at the municipality level and exploit the 1918–1920 influenza pandemic as a natural experiment to identify the impact of health shocks on pawnshop loans. Our main findings are as follows. First, the amount of pawnshop loans increased by 3.73–5.00% in the influenza pandemic years. This effect was similar in cities and counties. In addition, the redemption rate, which was approximately 90% in the 1910s and 1920s, was unchanged in the pandemic years. These results suggest that poor people used pawnshop loans to cope with adverse health shocks. Second, the pandemic increased average loan amounts, whereas the number of loans was not affected. This result indicates that those who usually relied on pawnshops borrowed intensively under a health shock, but others did not use pawnshops—even under an unexpected shock.

The present study contributes to the literature in the following three ways. First, it provides evidence that consumption was smoothed by pawnshop loans throughout Japan in both urban and rural areas. The natural experiment of the influenza pandemic even enables us to identify the response to an unexpected shock. Second, this study shows the role of financial institutions for poor people under an adverse health shock, using historical data from the prewar period. In development economics, there are a number of studies on risk sharing (e.g., Gertler and Gruber 2002; Lee and Sawada 2010; Liu 2016; Sawada and Takasaki 2017). Moreover, the growing body of the literature has revealed that microfinance helps households smooth consumption when health shocks occur (see Gertler et al., 2009; Islam and Maitra 2012). This study contributes to this stream of the development literature. Finally, in the context of Japanese financial history, it provides a general quantitative analysis on pawnshops in contrast to previous studies based on a small number of cases (e.g., Ioku and Shizume 2014; Saito 1989; Shibuya et al., 1982). Further, concerning the role of pawnshops, whereas the traditional view has emphasized the negative aspects (Shibuya et al., 1982), we shed light on the positive economic role for poor people.

## Pawnshops in Japan around 1920

In Japan, pawnshops, which date back to the twelfth or thirteenth century, were a traditional financial institution (Research Bureau of the Bank of Japan 1913, p. 87). The emergence of pawnshops in Japan was much earlier than in Europe, where the first pawnshop was established in Italy in 1462 (Research Bureau of the Bank of Japan 1913, p. 110). The number of pawnshops increased with the development of the market economy in Japan. In particular, they increased rapidly in the Meiji period (1868–1912) along with the growth of other modern financial institutions such as banks (Shibuya et al., 1982, pp. 342–343). As a result, pawnshops became prevalent nationwide in the early twentieth century. Although data encompassing all pawnshops are scarce, we can obtain data on the number of pawnshops and both the amount and the number of loans, redemption, and foreclosure in every prefecture in only 1924 from the Social Affairs Bureau (1926). This information is divided into two areas: city and county. Panel A of Table 1 shows several indices calculated based on the above data. It is noteworthy that there were so many pawnshops—17,807 in total, or 379 per prefecture. For comparison purposes, we present the numbers of other financial institutions, based on the Statistics Bureau of the Cabinet (1926), as well as pawnshops again in Panel B of Table 1. The number of pawnshops was 2.5 and 66.5 times greater than that of ordinary banks and mutual loan companies, respectively. Moreover, the number of pawnshops was twice as high as that of post offices that offered the postal savings system.[2] These facts highlight the ease of access to pawnshops. Indeed, the number of pawnshop loans was approximately 25 million, or 67,188 per day in 1924.

When a person utilized a pawnshop for the first time, they needed a referral from an existing customer. If the borrower did not receive a referral, in offering a loan, the borrower had to sign a contract, and the pawnshop checked his/her name and address to confirm that the goods to be pawned were not stolen or lost (Social Division of Osaka City 1926, pp. 15–16). By contrast, people with a referral could take out a loan without any contracts and checks. Thus, those who first applied for a loan usually brought a referral (Research Bureau of the

---

[2] James and Suto (2011) concluded that Japanese workers in the 1910s and 1920s saved easily because of the geographically widespread postal savings system. This conclusion emphasizes that access to pawnshops was easy.

Bank of Japan 1913, p. 91). In each case, pawnshops did not check borrowers' characteristics such as occupation and health status. Items to be pawned were one of the solutions to information asymmetry as it was easier for lenders to assess the value of an item to be pawned than to assess the financial status of a borrower (Kenttä 2016, p. 19). Hence, people could borrow money from pawnshops without any other requirements.

People could take out a loan within the total values of pawned items (Social Division of Osaka City 1926, p. 65). The value was assessed based on the prices in the secondhand market. Although the Act of Control of Pawnbroking was enacted in 1895, pawnshops usually followed traditions and customs rather than the Act. Thus, in accordance with the customs, they sold goods in foreclosure to antiquarians, whereas the pawnshops in other countries usually sold unredeemed goods by public auction. Moreover, the difference between the sales price of a pawned item and the amount of a defaulted loan became profit for the pawnshop because the ownership of the item belonged to the pawnshop after the foreclosure (Tokyo Pawnshop Association 1934). Therefore, pawnshops assessed the value of a pawned item so that they did not make a loss even if the item was unredeemed. If a borrower thought the assessed value was too low, he/she would utilize another pawnshop (Social Division of Osaka City 1926, p. 65). This competition between pawnshops prevented them from assessing the value of a pawned item as unreasonably low.

Pawnshops were the primary financial institutions for poor people in Japan—even in the 1920s. For instance, poor people visited pawnshops first when they needed money (Research Bureau of the Bank of Japan 1913, p. 86), many people associated pawnshops with commons (Tokyo Institute for Municipal Research 1925, p. 108), and pawnshops were the only financial institutions that helped low-income people (Tokyo Institute for Municipal Research 1926, p. 2). In fact, a survey of poor districts in Tokyo conducted in 1911 showed that 1,910 of 3,047 people (62.68%) had an existing loan from a pawnshop at the time of the survey (Regional Bureau of the Home Ministry 1912). There were various reasons other financial institutions were not preferred by poor people: banks required expensive collateral, usuries made excessive profits, while mutual loans were unreliable because of the vulnerability of participants and could not respond to participants' demand for money (Research Bureau of the Bank of Japan 1913; Shibuya 2001). Another survey of 15,634 poor households in Tokyo in 1933 indicated that pawnshop loans accounted for 69.68% of the total borrowing (Social Division of the Municipal Bureau of Academic Affairs of Tokyo 1935). This survey also showed that borrowing from relatives or friends only occurred 20.50% of the time. This means that poor people were more reliant on pawnshops than on their personal networks.

The reasons that pawnshops were popular are as follows. First, they accepted even inexpensive goods as pawns. Pawning items helped poor people, who could not access credit because they were usually considered to be unreliable, borrow money. Additionally, borrowers could avoid any trouble other than foreclosure if they did not redeem their pawn (Saito 1989, p. 5). In fact, informal finance accessed through networks based on kin, community, and workplace is not always good for the poor because of non-financial costs such as a lack of privacy (Collins et al., 2009). Therefore, it seems that a loan with an inexpensive pawn was preferable for the poor to that without any pawns. Shibuya et al. (1982) reported the numbers of pawns by their types based on the pawn ledgers of a pawnshop in January and June 1917. It is striking that clothes accounted for 88.97% of all pawns, whereas other types of pawns had much smaller shares: 2.77% were piece goods (*tanmono*), 1.24% mosquito nets, 1.18% Japanese futons, 1.06% accessories, and 0.48% watches. Obviously, most people borrowed money from pawnshops against their own clothes. Inexpensive clothes were pledgeable because Japan had a well-developed secondhand clothing market (Francks 2012).

Second, people could obtain small short-term loans flexibly. A borrower could take out a pawnshop loan quickly and redeem a pawned item by repaying the money at any time within the repayment term. Moreover, most small pawnshops were open until midnight (Social Division of Osaka City 1926, p. 67). Poor people often faced a shortage of living costs because of unexpected events such as illness, and thus their borrowing from pawnshops was mainly used for consumption (Shibuya 2000). Hence, it was important for the poor to be able to borrow as needed.

In addition, specific features of pawnshop loans, namely short term and small amounts, freed borrowers from a large interest burden. Since Japanese pawnshops did not intend to reduce poverty but were profit-oriented, their interest rates were not low. The maximum monthly interest regulated by the law depended on the loan amount; 4%, 3%, and 2.5% if the loan amount was under 1, 5, and 10 yen, respectively, and fixed at 0.01 yen if the amount was under 0.25 yen.[3] However, this interest rate seemed to matter little because borrowers repaid their loans before the interest ballooned. Indeed, Shibuya et al. (1982) provided an example that shows that 63.12% of loans from a pawnshop were repaid within a month even though pawnshop loans usually had repayment terms of six months.

## 1918–1920 influenza pandemic

The 1918–1920 influenza pandemic known as "Spanish flu" is one of the deadliest pandemics in history. During this period, 500 million people were infected and 50–100 million people died from influenza worldwide (Taubenberger and Morens 2006). Japan also experienced the pandemic with two waves whose peaks were in November 1918 and January 1920 (Ogasawara 2018b). According to the Sanitary Bureau of the Home Department (1922), which was the most comprehensive contemporary survey, there were 23,580,495 affected people (i.e., more than 40% of the population) and 385,029 influenza deaths in Japan between August 1918 and July 1920. However, this loss has been considered to be underestimated because of problems with determining cause of death and missing data. Therefore, Hayami (2006) and Richard et al. (2009) estimated the number of excess deaths between 1918 and 1920 as approximately 453,000 and 481,000, respectively. Their estimates suggest that the actual death toll associated with the 1918–1920 influenza pandemic was much larger than the reported death toll.[4]

---

[3] Pawnshops, especially large ones, did not necessarily apply the maximum interest rate. The Research Bureau of the Bank of Japan (1913) provides some examples in Tokyo. However, in addition to this legally regulated interest, more interest was required in accordance with customs (Shibuya et al. 1982).

One of the well-known features of the pandemic is the age distribution of mortality. Young adults experienced the highest risk of influenza death in 1918–1920, whereas influenza usually strikes the very young and the elderly (Gagnon et al., 2013; Langford 2002; Ogasawara 2017; Richard et al., 2009). This characteristic supports that the influenza pandemic caused an adverse health shock and a subsequent negative income shock to working people, as Karlsson et al. (2014) revealed that the pandemic led to a significant increase in poorhouse rates in Sweden.

Further, Spanish flu was unexpected, disappeared quickly, and affected a large proportion of the population at random (Karlsson et al., 2014). Owing to such characteristics, the pandemic has been widely used as a natural experiment in recent years (e.g., Acquah et al., 2017; Almond 2006; Brainerd and Siegler 2003; Lin and Liu 2014; Percoco 2016). In addition to the above advantages, the timing and intensity of the pandemic varied throughout Japan (Hayami 2006; Ogasawara 2017; Richard et al., 2009). Hence, we can use this regional heterogeneity for our estimation.

---

[4] Chandra (2013) estimated that the population loss including the decline in fertility was approximately two million during 1918–1919 in Japan. This result implies that the number of deaths related to the pandemic was even larger than the estimates of Hayami (2006) and Richard et al. (2009).

However, there is a paucity of available data on the influenza death toll at the county level. Therefore, we use the crude death rate, defined as the number of all deaths per 1,000 people, as a measure of the shock by the 1918–1920 influenza pandemic. Fig. 1 illustrates the average crude death rate and influenza death rate of prefectures where our observations are recorded, from 1915 to 1924. Obviously, these two death rates showed a similar trend during this period. In particular, not only the influenza death rate, but also the crude death rate rose dramatically in the pandemic years. This similarity allows us to use the crude death rate as a measure to exploit the exogenous pandemic in our regression analyses.

## Data and estimation strategy

Our main data sources are the statistical yearbooks annually published by each Japanese prefectural government (*Fuken tokeisho*). From these statistical yearbooks, we can obtain data on the population, death toll, and pawnshops at the city and county levels. Unfortunately, however, the data on death toll and pawnshops are only separately available for some prefectures. Therefore, we compiled panel data from 273 municipalities (i.e., 36 cities and 237 counties) in 19 prefectures between 1915 and 1924. Fig. 2 illustrates the spatial distribution of the prefectures where our sample cities and counties are recorded. Our observations are nationwide except in the Chubu region (central Japan). For comparison purposes, we calculated the amount of loans per pawnshop in 1924 based on our sample and the final column in Table 1, respectively; the results are 6,905 yen and 7,717 yen, respectively.

The latter is larger than the former, but it is mainly driven by Tokyo. Recalculated excluding Tokyo, the latter value becomes 6,696 yen, which is close to the former.

To investigate how many pawnshop loans were taken out to compensate for losses due to illness, we estimate the following fixed effects model. Our baseline model is given by

$$y_{it} = \alpha + \gamma CDR_{it} + \delta Pawnshop_{it} + \theta Population_{it} + \mathbf{x}'_{jt}\beta + v_i + \mu_t + \varepsilon_{it} \quad (1)$$

where $i$ indexes municipalities from 1 to 273, $j$ indexes prefectures from 1 to 19, and $t$ indexes years from 1915 to 1924. The dependent variable $y_{it}$ is the logarithm of the total amount of pawnshop loans, $CDR_{it}$ is the crude death rate (‰), $Pawnshop_{it}$ is the number of pawnshops per square kilometer, $Population_{it}$ is the logarithm of the population, $\mathbf{x}'_{jt}$ is a vector of the control variables at the prefecture level, and $\varepsilon_{it}$ is a random error term.[5] $v_i$ and $\mu_t$ represent the municipality and year fixed effects, respectively.

The variable *CDR* is of our interest. By taking advantage of the exogenous shock, namely the 1918–1920 influenza pandemic, we estimate the coefficient $\gamma$, which measures the effect of adverse health shocks on the loan amount. If this is positive and significant, the estimated coefficient indicates that pawnshops served as an intertemporal risk-coping mechanism.

Since we should consider the heterogeneity of the ease of access to pawnshops in terms of geographical proximity, the number of pawnshops per square kilometer is included in our estimation model as the measure.[6] The population indicates the number of possible

---

[5] In our regression analyses, all the variables that measure the amount of money are adjusted on the basis of the consumer price index provided by Okawa et al. (1967).

[6] Gertler et al. (2009) revealed that geographical proximity to financial institutions is

borrowers. Thus, a growing population was likely to lead to an increase in the loan amount.

As mentioned above, the health shock caused by the 1918–1920 influenza pandemic was exogenous. However, the crude death rate consists of not only deaths related to the pandemic but also unrelated deaths. This means that some factors in the error term might correlate with the unrelated deaths included in the crude death rate. To deal with this possible endogeneity, we employ the following control variables at the prefecture level. First, we control for the income level by using the logarithm of unit rice yields following Schneider and Ogasawara (2018). Although the pandemic struck randomly regardless of wealth level, some deaths not associated with the pandemic might be correlated to income. Hence, we exclude this possible correlation from our regression analyses. Second, the number of doctors per 100 people is included. This variable represents opportunities for consulting a doctor that are likely to decrease non-pandemic-related deaths in general (Inoue and Ogasawara 2018). These variables are our main controls. Third, we use the out-of-school rate (OOSR) for children of primary school age as a measure of poverty. Since the primary school enrollment rate in Japan was already 99.03% in 1920 (Archives Division of the Secretariat of the Ministry of Education 1925), the OOSR (i.e., 100 - the enrollment rate) seems to capture the proportion of the poorest households, which were generally unhealthier. Finally, we include the logarithm of the livestock population. This variable is interpreted as another measure of wealth. In addition to the aforementioned control variables, we use municipality fixed effects and year fixed effects in all specifications. These fixed effects control for all unobservable factors, including those constant over time, such as geographical characteristics, and those constant across municipalities, such as the impact of World War I. Our data sources for the

---

important for households to deal with adverse health shocks, using data in the 1990s in Indonesia.

control variables are described in Appendix.

## Effects of pandemic on pawnshop loans

Table 2 presents the estimation results of Eq. (1). Columns (1)–(3) show the estimates using only the municipality-level variables. In columns (4)–(6), we control for the influences of income and medical treatment. The results in columns (7)–(9) report the estimates from our baseline specification in which we include all the control variables. Further, there is a great difference in the amount of pawnshop loans between cities and counties, as shown in Panel A of Table 1. To consider this disparity, we conduct analyses using different samples. Columns (1), (4), and (7); (2), (5), and (8); and (3), (6), and (9) present our estimation results for both cities and counties, cities, and counties, respectively.

Columns (1)–(3) in Table 2 show that the crude death rate had a significantly positive effect on the amount of pawnshop loans. The estimated coefficients of *CDR* remain positive and significant even when we control for other factors to address the possible endogeneity, as shown in columns (4)–(9). These results are consistent with our hypothesis, suggesting that poor people used a pawnshop loan to cope with exogenous adverse health shocks. Moreover, the estimated coefficients of the crude death rate are stable in both cities or counties. Indeed, columns (2) and (3) report almost the same coefficients. Even after including the control variables, the magnitude in cities, which is 0.009 as shown in columns (5) and (8), is similar to that in counties shown in columns (6) and (9). Therefore, it is likely that there was no difference between cities and counties in terms of the increase in pawnshop loans attributed to the unexpected health shock.

The estimation results of the coefficients other than those of *CDR* are as follows (Table 2). The number of pawnshops per square kilometer is positively correlated to the loan amount only in cities. By contrast, the coefficients of population are not significant only in cities.

These results imply that the density of pawnshops was a binding constraint on loan amount only in cities, whereas population was a binding constraint on loan amount only in counties. We can interpret that cities had more potential borrowers than the population for which the existing pawnshops could provide loan services. As expected, the coefficients of rice yield are significantly negative. The negative coefficient of *Doctor* indicates that an increase in the opportunity to consult a doctor restrained borrowing from pawnshops. This result may indicate that the cost of consulting a doctor was lower than the cost of the decline in productivity caused by an adverse health shock.[7] The OOSR has a positive correlation with the pawnshop loan amount but this result is insignificant in cities. This finding is consistent with the fact that out-of-school children were concentrated in counties. No relationships are observed between the amount of loans and livestock population.

To examine the intensive and extensive margins, we conduct further estimates by Eq. (1) using the logarithm of the average loan amount (*AveLoan*) and the logarithm of the number of loans (*NumLoan*) as the dependent variable instead.[8] The estimated coefficients of *AveLoan* and *NumLoan* are interpreted as the effects on the intensive and extensive margins, respectively. Table 3 presents the estimated coefficients of *CDR* for both *AveLoan* in the second column and *NumLoan* in the third column. The main controls in the fourth column mean *Rice* and *Doctor* and the additional controls in the fifth column mean *OOSR* and *Livestock*. Panel A of Table 3 shows the results for the full sample. For the average loan amount, the estimation results are significantly positive and stable across the different specifications. This finding indicates that people borrowed more money than usual from pawnshops when they hit hard times due to the pandemic. The results for counties in Panel C are similar to the results in Panel A, while the results for cities in Panel B are positive but insignificant. Considering that the magnitudes of the estimated coefficients in Panel B are nearly equal to those in Panel C, it seems that the reason for the insignificance is the paucity of observations. By contrast, there are no relationships between the number of loans and crude death rate. This finding implies that only those who usually borrowed from pawnshops, who were likely to be poor, took out pawnshop loans to mitigate the adverse shock. In other words, our results suggest that a negative shock increased the intensive margin of pawnshop loans but had no impact on the extensive margin. Our possible explanation is that those who did not usually take out a pawnshop loan had savings, just in case (see James and Suto 2011). Hence, they did not need to rely on pawnshop loans.

---

[7] These two types of costs associated with illness are noted by Gertler and Gruber (2002) and Gertler et al. (2009).

[8] In this study, the intensive margin means the average loan amount and the extensive margin means the number of borrowers from pawnshops.

The results for *NumLoan* have another implication: regular borrowers did not increase their borrowing frequency. This finding is interpreted as rational behavior because the increase in the frequency of pawnshop loans required more total interest than the increase in the average loan amount if the total loan amounts were equal.

As described above, poor people took out larger loans from pawnshops than usual following the influenza pandemic. However, borrowers may not have redeemed their pawn when they took out a loan from a pawnshop to cope with an unexpected shock. In this case, our results mean that poor people used pawnshops to obtain money by selling their goods rather than through loans. To determine whether borrowers redeemed their pawn even when struck by the unforeseen influenza pandemic, we estimate the impact on the redemption rate by employing the same specification as in columns (7)–(9) in Table 2 except for the dependent variable. Table 4 shows the results for the proportion of redemption to the sum of redemption and foreclosure (%). Clearly, the estimated coefficients of the crude death rate are insignificant regardless of the sample used. Since the redemption rate around 1920 was approximately 90% (i.e., 89.20% as reported in Table 1 or 90.75%, which is the mean value in the full sample), this result is interpreted as evidence that most people who took out a pawnshop loan to address the health shock could redeem their pawn.

On the basis of some assumptions, we provide a more detailed interpretation of our estimation results as follows. The average crude death rates of our full sample in 1918, 1920, and at all other times are 26.84‰, 25.57‰, and 21.84‰, respectively. Therefore, the estimated coefficient in column (7) in Table 2 suggests that the amount of pawnshop loans increased by 5.00% in 1918 and 3.73% in 1920 because of the 1918–1920 influenza pandemic. Assuming that the total amount of loans that pawnshops would have made in 1918 and 1920 if the pandemic had not occurred was equal to the loan amount in 1924, namely 137,416,160 yen (see Table 1), it seems that 6,870,808 and 5,125,623 yen were borrowed to cover the losses caused by the adverse health shock. Further, under the assumption that the poverty rate was 9% at that time and only the poor among the 23,580,495 affected people borrowed more money from pawnshops than usual, the increase in victims' pawnshop loan amount per capita is calculated to be 5.65 yen.[9] In 1918 and 1920, the adjusted average wage per day of a female worker in the silk industry and a male day worker was 0.53 and 1.20 yen, respectively. Thus, our result suggests that they took out an additional pawnshop loan equal to 4.7–10.7 days' wages. This interpretation is likely to be consistent with the duration of influenza. The increased loan may be as much as they would have earned if they had not been affected. If poor people were more at risk of influenza infection than others, and assuming that they accounted for 20% of affected people, the increase in the pawnshop loan amount is calculated to be 2.54 yen per capita. That is, under this conservative assumption, our estimate implies that affected poor people increased their pawnshop loans by as much as 1.6–4.8 days' wages.

---

[9] Our assumption about the poverty rate is based on Ogasawara and Kobayashi (2015). They showed that the poverty headcount ratio was approximately 9% in the 1930s, and rising. Hence, the poverty rates in 1918 and 1920 might have been lower than 9%.

## Conclusion

In this study, we focused on pawnshop loans to examine a risk-coping strategy of poor people in prewar Japan. The natural experiment exploiting the 1918–1920 influenza pandemic enabled us to identify the response of pawnshop loans to an adverse health shock.

By using data on pawnshop loans in more than 250 municipalities from 1915 to 1924, we found that the loan amount increased by 3.73–5.00% because of the influenza pandemic. In addition, the redemption rate was stable at a high level regardless of the intensity of the shock. These findings suggest that people took out pawnshop loans to cope with the health shock. Our finding that regular users, who were generally from the low-income class, borrowed more money than usual following the unforeseen shock, indicates that pawnshop loans were a common risk-coping strategy for poor people in prewar Japan. However, our estimation results also indicate that those who had not usually taken out pawnshop loans did not rely on pawnshops even when the pandemic occurred. Thus, if they were relatively rich people and could smooth their consumption through interest-free methods such as savings, the difference in the risk-coping strategies that depended on income level exacerbated economic inequality, as Saaritsa (2011) remarked.

The results of this paper have the following implications on the literature. First, we found a new way for poor people to cope with risk. In previous studies in European countries, informal assistance through personal networks is regarded as a major coping strategy for risk. In this study, we show that a specific type of traditional financial institution—pawnshops— was widely used to cope with an unexpected adverse health shock in prewar Japan. This finding implies diversity in risk-coping strategies across time and space. Further, this study provides a useful insight to development studies, where the role of microfinance has been attracting a great deal of interest. Using the data from the years 1915 to 1924 in Japan, we discovered that private pawnshops served as community microfinance institutions. Finally, in the context of Japanese financial history, our findings suggest that revisions of the generally accepted view of the role of pawnshops are needed. In Japan, previous historical studies of pawnshops stressed that private pawnshops expanded poverty among small-scale producers through the repossession of pawned items (Shibuya et al., 1982, p. 353). In contrast, our quantitative analyses revealed that pawnshops actually functioned as a risk-coping device for poor people in Japan.

The spread of pawnshops as a risk-coping device is likely to relate to economic and financial conditions in prewar Japan. Before the 1930s, public financial services to poor people were limited. In fact, there were only one and five public pawnshops in 1918 and 1920, respectively (Tokyo Institute for Municipal Research 1926, pp. 164–168). This condition allowed private pawnshops to be prevalent nationwide without competition from public institutions. The profitability of private pawnshops was sufficiently high and thus did not need a support from the government. Since there were a large number of pawnshops, for example, more than 17,000 in 1924, they were easily accessible. In addition, their prevalence created price competition among pawnshops, which benefited borrowers through preventing unreasonable evaluation of the value of an item to be pawned (Research Bureau of the Bank of Japan 1913, p. 91). Thus, the prevalence of pawnshops was a condition by which poor people could utilize them for risk coping. Although formal systems of social insurance were underdeveloped in early twentieth-century Japan, pawnshops played an important role as a private social insurance system throughout the country.

———. "The long-run effects of pandemic influenza on the development of children from elite backgrounds: Evidence from industrializing Japan." *Economics and Human Biology* 31, (2018b): 125–37.

Ogasawara, Kota and Genya Kobayashi. "The impact of social workers on infant mortality in inter-war Tokyo: Bayesian dynamic panel quantile regression with endogenous variables." *Cliometrica* 9, no. 1 (2014): 97–130.

Ogilvie, Sheilagh, Markus Küpker, and Janine Maegraith. "Household debt in early modern Germany: Evidence from personal inventories." *Journal of Economic History* 72, no. 1 (2012): 134–67.

Okawa, Kazushi, Miyohei Shinohara, and Mataji Umemura. *Choki keizai tokei: Suikei to bunseki* [*Long-term economic statistics*], vol. 8. Tokyo: Toyo keizai sinposha, 1967.

Percoco, Marco. "Health Shocks and Human Capital Accumulation: The Case of Spanish Flu in Italian Regions." *Regional Studies* 50, no. 9 (2016): 1496–508.

Piketty, Thomas. *Capital in the Twenty-first Century*. Cambridge, MA: Belknap Press of Harvard University, 2014.

Regional Bureau of the Home Ministry. *Saimin chosa tokeihyo* [*Statistical survey of lower income class*]. Tokyo, 1912.

Research Bureau of the Bank of Japan. *Shichiya ni kansuru chosa* [*Survey on pawnbroking*]. Tokyo, 1913.

Richard, Stephanie A., N. Sugaya, L. Simonsen, M. A. Miller, and C. Viboud. "A comparative study of the 1918–1920 influenza pandemic in Japan, USA and UK: Mortality impact and implications for pandemic planning." *Epidemiology and Infection* 137, no. 8 (2009): 1062–72.

Saaritsa, Sakari. "Informal transfers, men, women and children: Family economy and informal social security in early 20th century Finnish households." *The History of the Family*

13, no. 3 (2008): 315–31.

———. "The poverty of solidarity: The size and structure of informal income smoothing among worker households in Helsinki, 1928." *Scandinavian Economic History Review* 59, no. 2 (2011): 102–27.

Saito, Hiroshi. *Shichiyashi no kenkyu* [*Study on history of pawnbroking*]. Tokyo: Shinhyoron, 1989.

Sanitary Bureau of the Home Department. *Ryukosei kanbo* [*Influenza pandemic*]. Tokyo, 1922.

Sawada, Yasuyuki and Yoshito Takasaki. "Natural Disaster, Poverty, and Development: An Introduction." *World Development* 94, (2017): 2–15.

Schneider, Eric B. and Kota Ogasawara. "Disease and child growth in industrialising Japan: Critical windows and the growth pattern, 1917–39." *Explorations in Economic History* 69, (2018): 64–80.

Scott, Peter M. and James Walker. "Working-class household consumption smoothing in interwar Britain." *Journal of Economic History* 72, no. 3 (2012): 797–825.

Shibuya, Ryuichi. *Korigashi kinyu no tenkai kozo* [*Expansion of usuries*]. Tokyo: Nihon tosho senta, 2000.

———. *Shomin kinyu no tenkai to sesaku taio* [*Expansion of financial institutions for poor people and policy response*]. Tokyo: Nihon tosho senta, 2001.

Shibuya, Ryuichi, Kameji Suzuki, and Shojiro Ishiyama. *Nihon no shichiya* [Pawnshop in Japan]. Tokyo: Waseda daigaku shuppanbu, 1982.

Social Affairs Bureau. *Koeki shichiko no genzei* [*Current situation of public pawnshops*]. Tokyo, 1926.

Social Division of the Municipal Bureau of Academic Affairs of Tokyo. *Saimin kinyu ni kansuru chosa* [*Survey on financial institutions for poor people*]. Tokyo, 1935.
23

# Tables

## Table 1: Pawnshops and other financial institutions in 1924

|  | Average value of prefectures | | | Total |
|---|---|---|---|---|
|  | All | City area | County area |  |
| Panel A: Pawnshop |  |  |  |  |
| Number of pawnshops | 378.87 | 122.11 | 256.64 | 17,807 |
| Amount of loans (yen) | 2,923,748.13 | 1,865,761.57 | 1,057,900.26 | 137,416,160 |
| Number of loans | 521,782.15 | 313,708.36 | 208,062.87 | 24,523,760 |
| Average loan amount (yen) | 5.37 | 5.87 | 5.16 | - |
| Redemption rate (%) | 89.20 | 88.53 | 89.77 | - |
|  |  |  |  |  |
| Panel B: Number of financial institutions |  |  |  |  |
| Pawnshop | 378.87 | 122.11 | 256.64 | 17,807 |
| Ordinary bank | 147.94 | - | - | 6,953 |
| Mutual loan company | 5.70 | - | - | 268 |
| Post office | 183.68 | - | - | 8,633 |

*Notes*: The average loan amount is calculated by dividing the amount of loans by the number of loans. The redemption rate is defined as the proportion of redemption to the sum of redemption and foreclosure.

*Sources*: See the text.

Table 2: Responses of loan amount to the influenza pandemic

|  | (1) Full | (2) City | (3) County | (4) Full | (5) City | (6) County | (7) Full | (8) City | (9) County |
|---|---|---|---|---|---|---|---|---|---|
| CDR | 0.013*** | 0.012* | 0.014*** | 0.011** | 0.009* | 0.012** | 0.010** | 0.009* | 0.011** |
|  | (0.005) | (0.006) | (0.005) | (0.005) | (0.005) | (0.005) | (0.004) | (0.005) | (0.005) |
| Pawnshop | 0.000 | 0.002 | -0.001 | 0.000 | 0.002* | -0.002 | -0.000 | 0.002* | 0.002 |
|  | (0.002) | (0.001) | (0.009) | (0.002) | (0.001) | (0.009) | (0.002) | (0.001) | (0.009) |
| Population | 2.371*** | 0.270 | 2.805*** | 2.321*** | 0.130 | 2.769*** | 2.345*** | 0.138 | 2.797*** |
|  | (0.279) | (0.293) | (0.310) | (0.286) | (0.291) | (0.314) | (0.289) | (0.242) | (0.314) |
| Rice |  |  |  | -0.346*** | -0.479** | -0.331*** | -0.269*** | -0.497** | -0.231** |
|  |  |  |  | (0.097) | (0.201) | (0.108) | (0.091) | (0.214) | (0.101) |
| Doctor |  |  |  | -2.222** | -6.077*** | -1.751 | -2.816** | -5.799*** | -2.419** |
|  |  |  |  | (1.079) | (2.072) | (1.139) | (1.092) | (1.963) | (1.171) |
| OOSR |  |  |  |  |  |  | 0.093*** | -0.031 | 0.118*** |
|  |  |  |  |  |  |  | (0.029) | (0.112) | (0.028) |
| Livestock |  |  |  |  |  |  | 0.199 | -0.109 | 0.199 |
|  |  |  |  |  |  |  | (0.246) | (0.243) | (0.279) |
| Municipality FE | Yes | Yes | Yes | Yes | Yes | Yes | Yes | Yes | Yes |
| Year FE | Yes | Yes | Yes | Yes | Yes | Yes | Yes | Yes | Yes |
| $F$-statistics | 24.91*** | 2.12 | 33.75*** | 23.85*** | 9.62*** | 29.07*** | 19.24*** | 8.22*** | 25.01*** |
| $R$-squared | 0.4103 | 0.2103 | 0.4478 | 0.4151 | 0.2547 | 0.4515 | 0.4226 | 0.2566 | 0.4626 |
| Observations | 2,306 | 291 | 2,015 | 2,306 | 291 | 2,015 | 2,306 | 291 | 2,015 |

* = Significant at the 10 percent level.

** = Significant at the 5 percent level.

*** = Significant at the 1 percent level.

*Notes*: The dependent variable is the logarithm of the total amount of pawnshop loans. Standard errors clustered at the municipality level are in parentheses.

*Sources*: See the text and Appendix.

Table 3: Responses of the intensive and extensive margins

| | Dependent variable | | Main controls | Additional controls | Municipality- and Year- FE | Number of observations |
|---|---|---|---|---|---|---|
| | *AveLoan* | *NumLoan* | | | | |
| Panel A: Full sample | 0.008** | 0.003 | No | No | Yes | 1,986 |
| | (0.003) | (0.004) | | | | |
| | 0.007** | 0.002 | Yes | No | Yes | 1,986 |
| | (0.003) | (0.004) | | | | |
| | 0.007** | 0.001 | Yes | Yes | Yes | 1,986 |
| | (0.003) | (0.004) | | | | |
| Panel B: City | 0.007 | 0.005 | No | No | Yes | 233 |
| | (0.008) | (0.008) | | | | |
| | 0.007 | 0.003 | Yes | No | Yes | 233 |
| | (0.008) | (0.008) | | | | |
| | 0.007 | 0.003 | Yes | Yes | Yes | 233 |
| | (0.008) | (0.008) | | | | |
| Panel C: County | 0.008** | 0.004 | No | No | Yes | 1,753 |
| | (0.003) | (0.004) | | | | |
| | 0.008** | 0.002 | Yes | No | Yes | 1,753 |
| | (0.004) | (0.004) | | | | |
| | 0.007** | 0.001 | Yes | Yes | Yes | 1,753 |
| | (0.003) | (0.005) | | | | |

\* = Significant at the 10 percent level.

\*\* = Significant at the 5 percent level.

\*\*\* = Significant at the 1 percent level.

*Notes*: The second and third columns present the estimated coefficients of *CDR* and its standard errors clustered at the municipality level in parentheses.

*Sources*: See the text and Appendix.

Table 4: Impacts of the influenza pandemic on the redemption rate

|  | (1) Full | (2) City | (3) County |
|---|---|---|---|
| *CDR* | 0.041 | -0.019 | 0.047 |
|  | (0.065) | (0.093) | (0.077) |
| *Pawnshop* | 0.017 | 0.020 | 0.175 |
|  | (0.013) | (0.016) | (0.118) |
| *Population* | -6.421* | -5.892* | -7.735* |
|  | (3.315) | (3.430) | (4.426) |
| *Rice* | 1.637 | 1.919 | 1.610 |
|  | (2.803) | (2.224) | (3.216) |
| *Doctor* | 5.858 | 7.308 | 7.738 |
|  | (16.656) | (26.954) | (19.063) |
| *OOSR* | -0.525 | -1.047* | -0.454 |
|  | (0.364) | (0.537) | (0.402) |
| *Livestock* | 10.453*** | 3.459 | 12.012*** |
|  | (3.055) | (5.226) | (3.431) |
| Municipality FE | Yes | Yes | Yes |
| Year FE | Yes | Yes | Yes |
| *F*-statistics | 3.44*** | 2.65** | 3.04*** |
| *R*-squared | 0.1678 | 0.2864 | 0.1630 |
| Observations | 2,306 | 291 | 2,015 |

\* = Significant at the 10 percent level.

\*\* = Significant at the 5 percent level.

\*\*\* = Significant at the 1 percent level.

*Notes*: The dependent variable is the redemption rate (%). Standard errors clustered at the municipality level are in parentheses.

*Sources*: See the text and Appendix.

# Figures

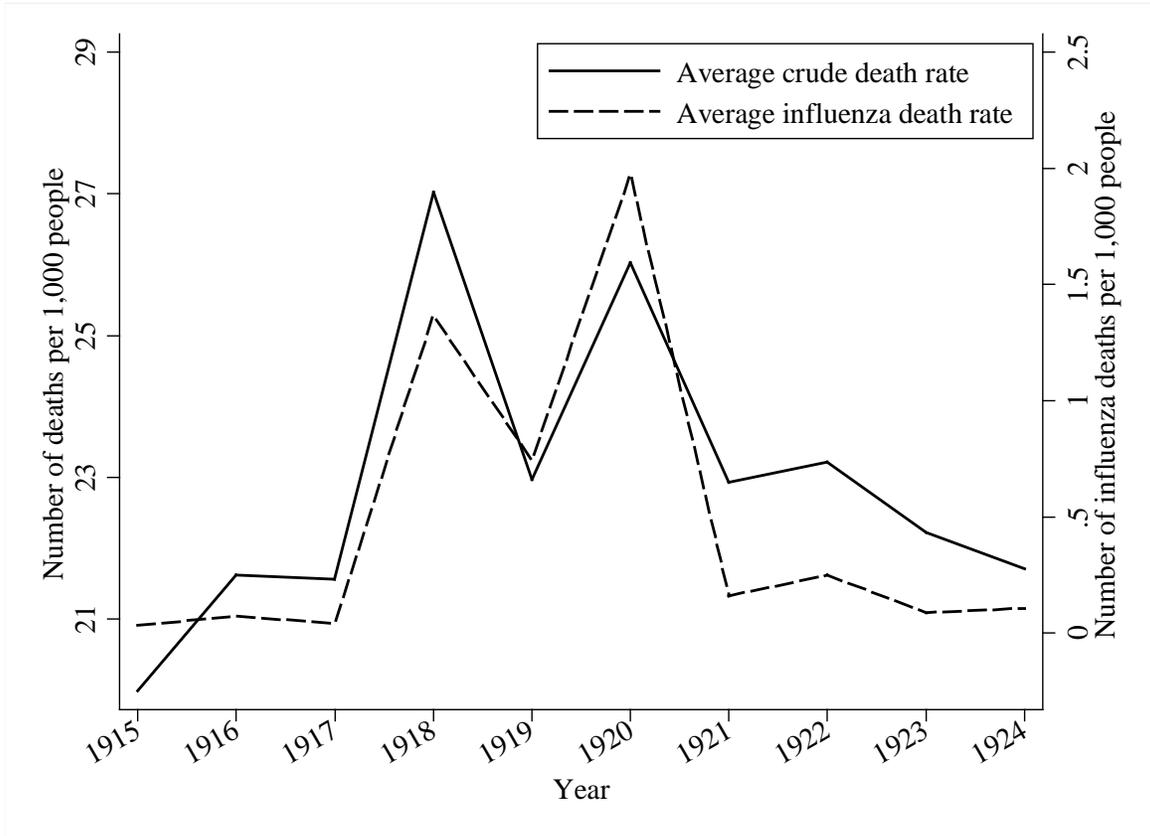

Figure 1: Average crude death rate and influenza death rate, 1915–1924

*Notes*: The crude death rate and influenza death rate are the average values of prefectures where our observations are recorded.

*Sources*: Statistics Bureau of the Cabinet (1918–1925).

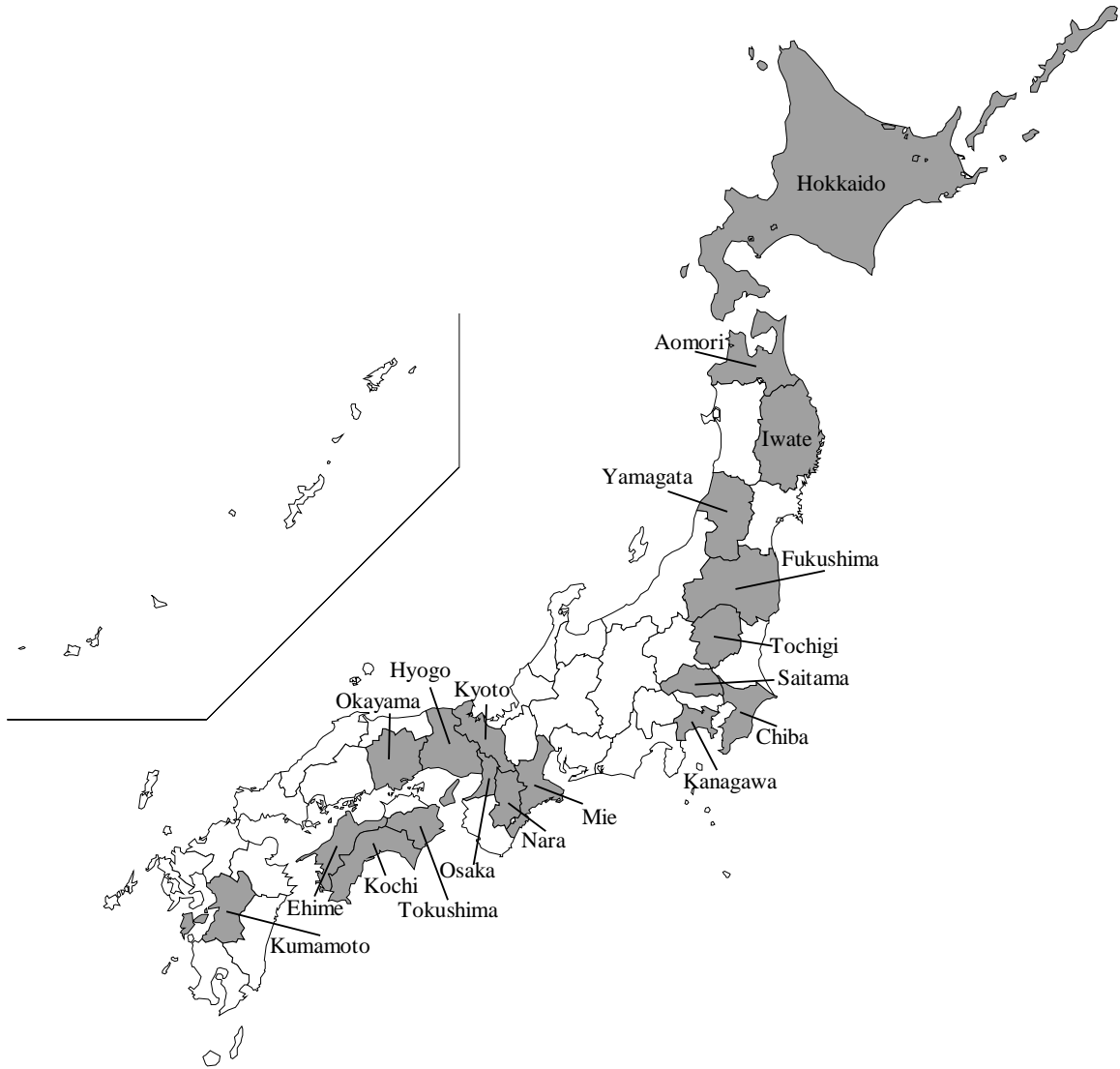

Figure 2: Spatial distribution of our observations

# Appendix

## Data sources

Our main data sources are *Fuken tokeisho*, which are statistical yearbooks annually published by each Japanese prefectural government, in 1915–1924 of 19 prefectures: Hokkaido, Aomori, Iwate, Yamagata, Fukushima, Tochigi, Saitama, Chiba, Yokohama, Mie, Kyoto, Osaka, Hyogo, Nara, Shimane, Tokushima, Ehime, Kochi, and Kumamoto. The data for each prefecture are available for 1918, 1920, and more than one other year. From the statistical yearbooks, we obtained the municipality level data, namely, the number of pawnshops, the total amount of loans, the frequency of loans, redemption, and foreclosure, the number of deaths, and population. Since the population census is conducted every five years in Japan, the data on population for non-census years is interpolated. The data from *Fuken tokeisho* is used for making the independent variables and the key dependent variables.

In terms of control variables, we compiled prefecture level data based on governmental statistical materials as follows. *Nihon teikoku tokei nenkan*, that is, Japanese statistical yearbooks are our source for a rice yield, land area planted with rice, and the number of doctors. We calculated the rice yield per unit by dividing the rice yield by the land area. The OOSR for children of primary school age is described in *Nihon teikoku monbusho nenpo*. The material consisting of vol. 1 and vol. 2 was published annually and recorded information about education and religion. For livestock population, we used *Noji tokei* and *Noji tokei hyo* where statistical agricultural information is described. The definition of the livestock population is the number of cows and horses for cultivation. For variables used in the section "Effects of pawnshop loans", we obtained the data from the census (*Kokusei chosa hokoku*) for 1920.

## Statistical materials